\RequirePackage[2020-02-02]{latexrelease}
\documentclass[twocolumn]{aastex631}
\usepackage{amssymb}
\usepackage{soul}
\usepackage{natbib}
\usepackage{multirow}
\usepackage{graphicx}
\usepackage{tabularx}
\usepackage{verbatim}
\usepackage{color}
\usepackage{ulem}
\usepackage{subfigure}
\usepackage{booktabs,tikz}
\makeatletter
\global\let\tikz@ensure@dollar@catcode=\relax
\makeatother
\usepackage{urwchancal}

\newcommand\soutm{\bgroup\markoverwith
{\textcolor{black}{\rule[0.5ex]{2pt}{0.8pt}}}\ULon}

\shorttitle{Connection between the Intracluster Light and its Host Halo}
\shortauthors{Contini et al.}

\begin{document}

\title{The Connection between the Intracluster Light and its Host Halo: Formation Time and Contribution from Different Channels}

\correspondingauthor{Emanuele Contini}
\email{emanuele.contini82@gmail.com}
\author{Emanuele Contini}
\affil{Department of Astronomy and Yonsei University Observatory, Yonsei University, 50 Yonsei-ro, Seodaemun-gu, Seoul 03722, Republic of Korea}
\author{Jinsu Rhee}
\affil{Department of Astronomy and Yonsei University Observatory, Yonsei University, 50 Yonsei-ro, Seodaemun-gu, Seoul 03722, Republic of Korea}
\affil{Korea Astronomy and Space Science Institute, 776, Daedeokdae-ro, Yuseong-gu, Daejeon 34055, Republic of Korea}
\author{San Han}
\affil{Department of Astronomy and Yonsei University Observatory, Yonsei University, 50 Yonsei-ro, Seodaemun-gu, Seoul 03722, Republic of Korea}
\author{Seyoung Jeon}
\affil{Department of Astronomy and Yonsei University Observatory, Yonsei University, 50 Yonsei-ro, Seodaemun-gu, Seoul 03722, Republic of Korea}
\author{Sukyoung K. Yi}
\affil{Department of Astronomy and Yonsei University Observatory, Yonsei University, 50 Yonsei-ro, Seodaemun-gu, Seoul 03722, Republic of Korea}

\begin{abstract}
We extend the analysis presented in \cite{contini2023a} to
higher redshifts, up to $z=2$, by focusing on the relation between the intracluster light (ICL) fraction and the halo mass, its dependence with redshift, role played by the halo concentration and formation time, in a large sample of simulated galaxy groups/clusters with $13\lesssim  \log M_{halo} \lesssim 15$. Moreover, a key focus is to isolate the relative contributions provided by the main channels for the ICL formation to the total amount.
The ICL fraction at higher redshift is weakly dependent on halo mass, and comparable with that at the present time, in agreement with recent observations. Stellar stripping, mergers and pre-processing are the major responsible channels of the ICL formation, with stellar stripping that accounts for $\sim 90\%$ of the total ICL, regardless of halo mass and redshift. Pre-processing is an important process for clusters to accrete already formed ICL.
The diffuse component forms very early, $z\sim 0.6$, and its formation depends on both concentration and formation time of the halo, with more concentrated and earlier formed haloes that assemble their ICL earlier than later formed ones. The efficiency of this process is independent of halo mass, but increases with decreasing redshift, which implies that stellar stripping becomes more important with time as the concentration increases. This highlights the link between the ICL and the dynamical state of a halo: groups/clusters that have a higher fraction of diffuse light are more concentrated, relaxed and in an advanced stage of growth.
\end{abstract}

\keywords{galaxies: clusters: general (584) galaxies: formation (595) --- galaxies: evolution (594) --- methods: numerical (1965)}

\section{Introduction}
\label{sec:intro}

The intra-cluster light (ICL) is a diffuse light that permeates the medium between galaxies in groups and clusters (\citealt{gonzalez2013,contini2014,mihos2017,iodice2017,tang2018} and many others). Due to its diffuse nature, the ICL is made of stars that are not bound to any galaxy in the host halo, but feel only the potential well of the dark matter halo. Several studies (see the reviews of \citealt{contini2021,montes2022,arnaboldi2022} and reference therein) mainly in the last two decades focused on the origin of the ICL, that is, on the mechanisms that bring to its formation, the main properties and its link with the overall dynamical evolution of the host group/cluster.

The ICL is a faint component, usually associated with the brightest group/cluster galaxy (BGG/BCG). Given its faint nature (even below the level of the sky), it is not trivial to detect it (e.g., \citealt{montes2014}), in particular at higher redshift. Once the contribution of the sky is removed, the primary difficulty is to separate it from the satellite galaxies (e.g., \citealt{presotto2014}), by masking the light coming from them. One more task would be that of separating it from the light coming from the BGG/BCG (hereafter BCG, \citealt{burke2015,iodice2017,demaio2018,spavone2018,ragusa2022} and others). A full separation of the two components will never be possible, considering the fact that they are embedded. Nevertheless, there are several techniques to isolate the ICL from the BCG, with pros and cons, like a surface brightness cut (e.g., \citealt{zibetti2005}), a profile fitting (e.g., \citealt{zhang2019}), or a simple distance (usually called transition radius) from the center of the BCG+ICL system that can minimize the lost of light from both components (e.g., \citealt{montes2018}). It must be noted, however, that many authors (e.g., \citealt{werner2023}) do not attempt to separate the two components and study them as a unique system, BCG+ICL.

Studying the properties of the ICL, both from the observational and theoretical points of view, can shed some light on the mechanisms that brought to its formation. In \cite{contini2023a} (hereafter C23a), we have discussed all the possible mechanisms invoked in the literature, and stated that only three of them are important: stellar stripping of satellite galaxies (\citealt{rudick2009,rudick2011,martel2012,contini2014,contini2018,demaio2015,demaio2018,montes2018}), mergers between satellites and the BCG (\citealt{monaco2006,murante2007,contini2014,burke2015,groenewald2017,joo2023}), and pre-processing/accretion of ICL from other haloes (\citealt{mihos2005,sommer-larsen2006,contini2014,ragusa2023}). The first two are direct ways to form ICL, because it is formed within the virial radius of the host halo, while pre-processing/accretion is a mechanism through which the ICL is incorporated in a given object, but it was formed elsewhere and via the other two mechanisms (see a full discussion of this point in C23a).

In C23a we focused on the link between the ICL and the dynamical state of the host halo, by exploring the role of the concentration (assuming an NFW profile, \citealt{nfw1997}) in the formation of the ICL. We found that, on average, more concentrated haloes have a larger fraction of ICL, which means that the processes responsible for its formation are also dependent on the concentration of a given object. We then suggested the concentration to be the main driver of the formation of the ICL. Observationally (e.g., \citealt{furnell2021}), the ICL fraction in the local universe does not seem to depend much on the halo mass, in line with many theoretical expectations (e.g., \citealt{contini2014}), and the dependence with redshift is still not clear (e.g., \citealt{montes2019b}). Only recently, \cite{joo2023} found that the ICL fraction of around 10 objects at redshift $1.0<z<1.8$ is similar to that in the local universe, underlining the fact that the contribution of the ICL to the total light of those objects is already important at high redshift. If this is true, it means that the overall growth of the stellar mass (BCG included) goes in parallel with that of the ICL.

In this paper (the second of a series of three) we want to address three main points, by approaching them in the same way it was done in C23a, i.e., through a semi-analytic model (SAM). The first main goal is to investigate the fractional budget of ICL in a wide range of galaxy groups and clusters \footnote{The sample comprises 2509 haloes in the mass range $13.0 \leq \log M_{200} \leq 15.0$ at $z=0$, which contain at least four satellites with $\log M_* \geq 8.5$ within the virial radius $R_{200}$.} (same sample used in C23a) as a function of redshift, and see whether or not our model predictions are comparable with observational measurements and if there is any trend, increasing or decreasing ICL fraction, with redshift. The second goal is to isolate the contribution of the different channels for the formation of the ICL, as a function of both halo mass and redshift. Hence, we are going to probe on which percentages stellar stripping and mergers (the direct channels) contribute to the total ICL at different redshift, from the local universe up to $z=2$, and how important pre-processing (indirect channel) is as a function of halo mass and redshift. Considering the results found in C23a, the third and last goal is to prove that there is a link between the formation of the ICL and that of its host halo. The aim is to strengthen former conclusions that the formation and evolution of the ICL is closely related to the evolutionary history of its host halo, so that the ICL can be a reliable candidate for studying the dynamical state of dark matter haloes (\citealt{zibetti2005,jee2010,giallongo2015,harris2017,montes2019b,asensio2020,sampaio2021,contini2020b,contini2021b,kluge2021,deason2021,contini2022,yoo2022}).

The structure of the paper is as follows. In Section 2 we briefly describe the set of simulations used and summarize the key points of the prescriptions for the formation of the ICL in our SAM. A detailed description of both simulations and model can be found in C23a. In Section 3 we show our analysis and briefly discuss the main results, while a full discussions of them is presented in Section 4. Finally, in Section 5 we summarize our main conclusions. As in C23a, stellar masses are computed with the assumption of a \cite{chabrier2003}
initial mass function and, unless otherwise stated, all units are h-corrected.

\section[]{Methods}
\label{sec:methods}

In order to carry out the analysis presented in Section 3, we take advantage of the same set of simulations used in C23a, with the exception of YS25HR and YS50HR. The set of simulations were performed by using {\small GADGET-4} (\citealt{springel2021}), the latest version of the {\small GADGET}-code. The detailed information of the set of simulations can be found in C23a, while here we
just summarize the main features.

The set is characterized by four simulations, i.e., four different boxes spanning a volume that goes from $(50 Mpc/h)^3$ up to $(200 Mpc/h)^3$, and with the same softening length, 3 $kpc/h$. Moreover, the simulations share the same following Planck 2018 cosmology (\citealt{planck2020}): $\Omega_m=0.31$ for the total matter density, $\Omega_{\Lambda}=0.69$ for the cosmological constant, $n_s=0.97$ for the primordial spectral index, $\sigma_8=0.81$ for the power spectrum normalization, and $h=0.68$ for the normalized Hubble parameter. The simulations ran from $z=63$ to $z=0$ and the data have been stored in 100 discrete snapshots from $z=20$ to $z=0$, allowing us to have a good refinement, which is important in the context of SAMs for prescriptions such as star formation and tidal stripping.
Tidal stripping is one of the two channels implemented in the model and responsible for the formation of the ICL. Another prescription that takes into account the formation of the diffuse light is implemented in the model any time a merger between a satellite and the BCG occurs. For the sake of simplicity, here we briefly summarize the mean features of both prescriptions, and we refer the reader to C23a (and references therein) for further details.

The possible channels for the formation of the ICL in our SAM are tidal stripping, mergers and pre-processing. As mentioned above, the first two are direct ways to form ICL, while pre-processing or accretion is an indirect way, in the sense that the ICL was already formed elsewhere before being accreted by a given halo.

{\small {\bf TIDAL STRIPPING}}: conceptually speaking this channel is very trivial, in the sense that it considers the tidal forces acting between the potential well of the dark matter halo and satellites orbiting around its centre. At each snapshot, the SAM derives the tidal radius of the satellites and compares it with their size, either bulge or disk radii. If the tidal radius is smaller than the radius of the bulge, the satellite is completely destroyed, but, if the tidal radius lies between the bulge and the edge of the disk, only the stellar mass outside the limit given by the tidal radius is removed from the disk. In both cases, the stellar mass that has been stripped ends up in the ICL component associated to the current BCG. It must be noted that there are two kinds of satellites in our SAM: those that still keep their
own subhalo (type1), and those that have lost it (type2 or orphans). The prescription as described is directly applied to type2, while for type1 satellites the model requires the extra condition that the half mass radius of the dark matter subhalo has to be smaller than the half mass radius of the disk. This assures that the subhalo has lost a significant amount of mass before stellar stripping can begin. If the condition is met, the given satellite loses also the ICL that was associated with it.

{\small {\bf MERGERS}}: this prescription is very simple. Every time a merger, minor or major, between a satellite and the current BCG occurs, 20\% of the mass of the satellite is moved to the ICL component. A full discussion of the reasons for the percentage chosen can be found in \cite{contini2014}.  It was the result of control simulations of galaxy groups (\citealt{villalobos2012}), from which we derived that the typical fraction of stellar mass that gets unbound during mergers peaks on 0.2, and we extended its validity to all scales. Later works (e.g. \citealt{contini2018,contini2019}) confirmed the validity of the assumption on larger scales.

{\small {\bf PRE-PROCESSING}}: as mentioned above and discussed in C23a, this channel is not a direct way to produce ICL. Indeed, the ICL is already formed via one of the channels described above, and later accreted by a given halo. However, pre-processing can be an important channel for increasing the amount of ICL during the assembly of groups and clusters. In our SAM, pre-processed ICL is given by the accretion of the ICL associated to those galaxies that have been accreted during a merger between haloes. In fact, when two haloes merge, one of them keeps being the main halo, and all the galaxies associated with the other one are accreted. These galaxies then become satellites of the new halo formed after the merger. If, among them, some become type2, they will lose the ICL associated with them which will be moved to that associated with the current BCG. Among type1 satellites, if they experience an episode of stripping (as described above), their ICL component will have the same fate, i.e., it will move to the ICL associated with the BCG. Ergo, pre-processed ICL is given by the ICL associated with galaxies that have been accreted, but it was formed in other haloes.

To summarize, the total ICL that a group or cluster can account for is formed either by stellar stripping or mergers. Some of it can be formed outside the halo and then being accreted via the mergers that
the group or cluster experienced during its assembly.

\section{Results}
\label{sec:results}

\begin{figure*}
\centering
\includegraphics[width=0.73\textwidth]{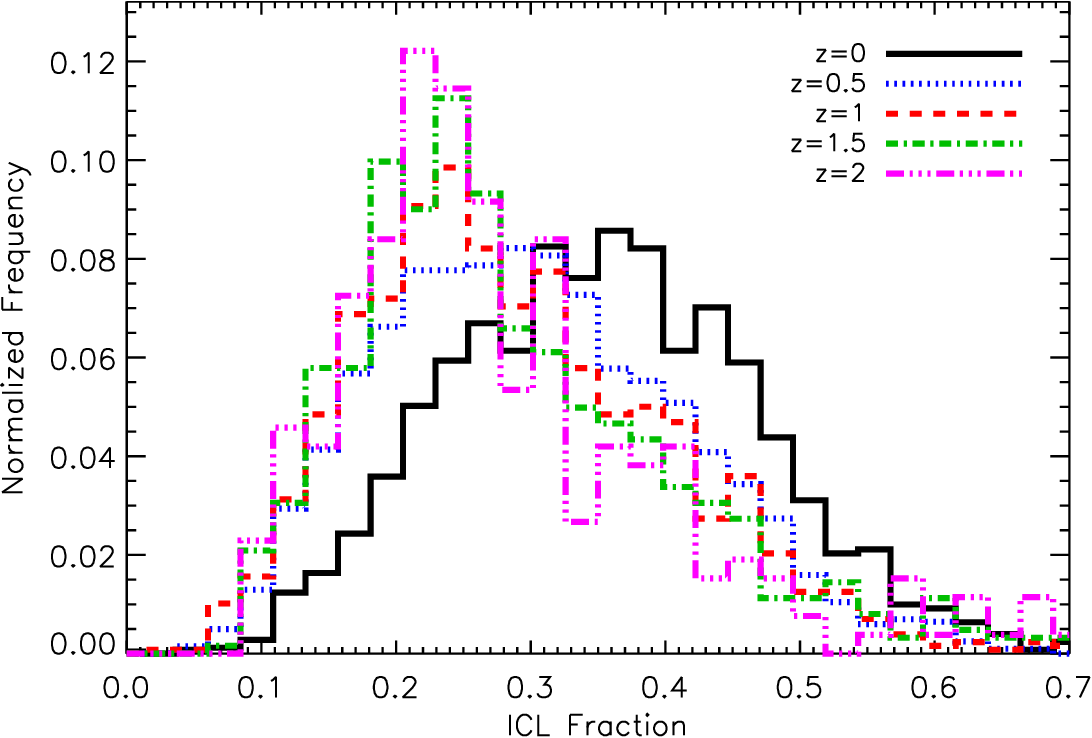}
\caption{Distributions of the ICL fraction in our samples of haloes at different redshifts, from $z=2$ (magenta line) down to $z=0$ (black line). As found in C23a, the distributions can be approximated with a gaussian at all
redshift, highlighting the fact that the processes responsible for the ICL formation are \textit{stochastic}. The interesting message of the plot is that the peak of the distribution tends to increase from higher
to lower redshift, being around 0.23 at $z=2$ and reaching about 0.35 at the present time. This indicates that the fraction of ICL gradually increases as time goes by, meaning that the processes for the ICL formation become more and more important with time, but the majority of the formation happens after $z=1$, and in particular between $z=0.5$ (blue line) and the present time (black line).}
\label{fig:histo}
\end{figure*}

In Section 1 we anticipated the three main goals of this study, namely: (a) to probe the ICL fraction in groups and clusters as a function of redshift; (b) to isolate the contribution coming from the different channels, direct and indirect; (c) to investigate the link between the formation of the ICL and that of its host halo. Before starting the analysis, it must be noted that some of the results shown below are dependent on the particular definitions we have chosen, such as those regarding the redshifts of formation of ICL and host halo. We state here, and remind the reader below where necessary, that we have also tried similar definitions. In all cases, the overall pictures do not change, even though quantitatively speaking some of the results change due to the different definitions. We will clarify every single case at the due time.

\begin{figure*}
\begin{center}
\begin{tabular}{cccc}
\includegraphics[scale=.48]{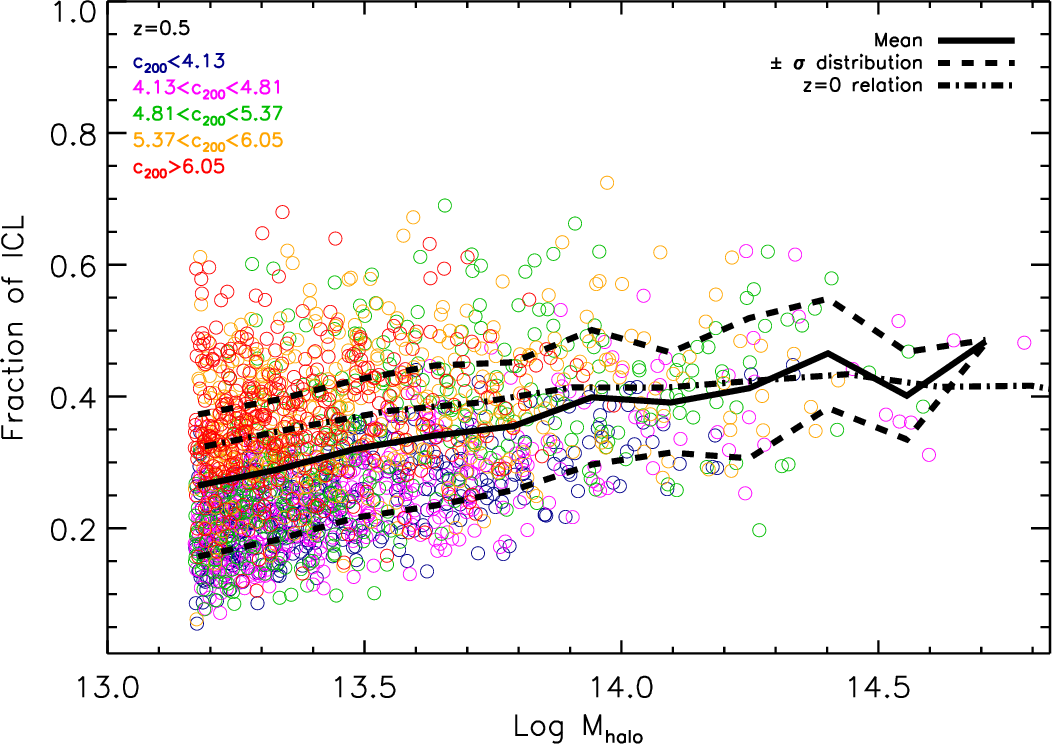} &
\includegraphics[scale=.48]{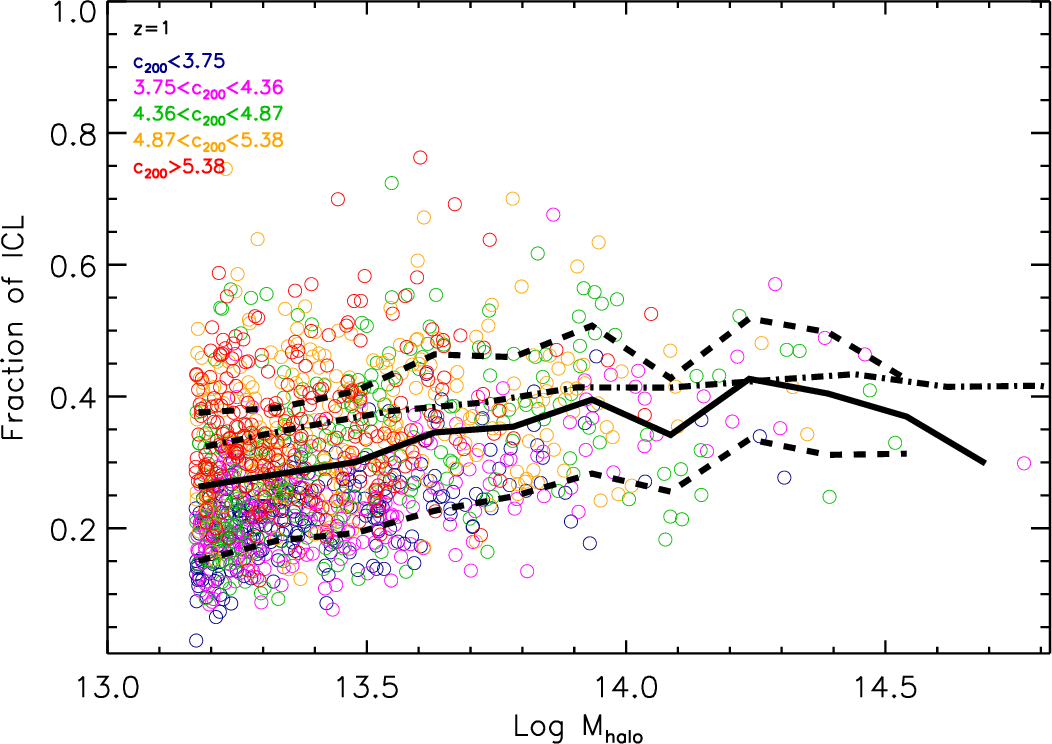} \\
\includegraphics[scale=.48]{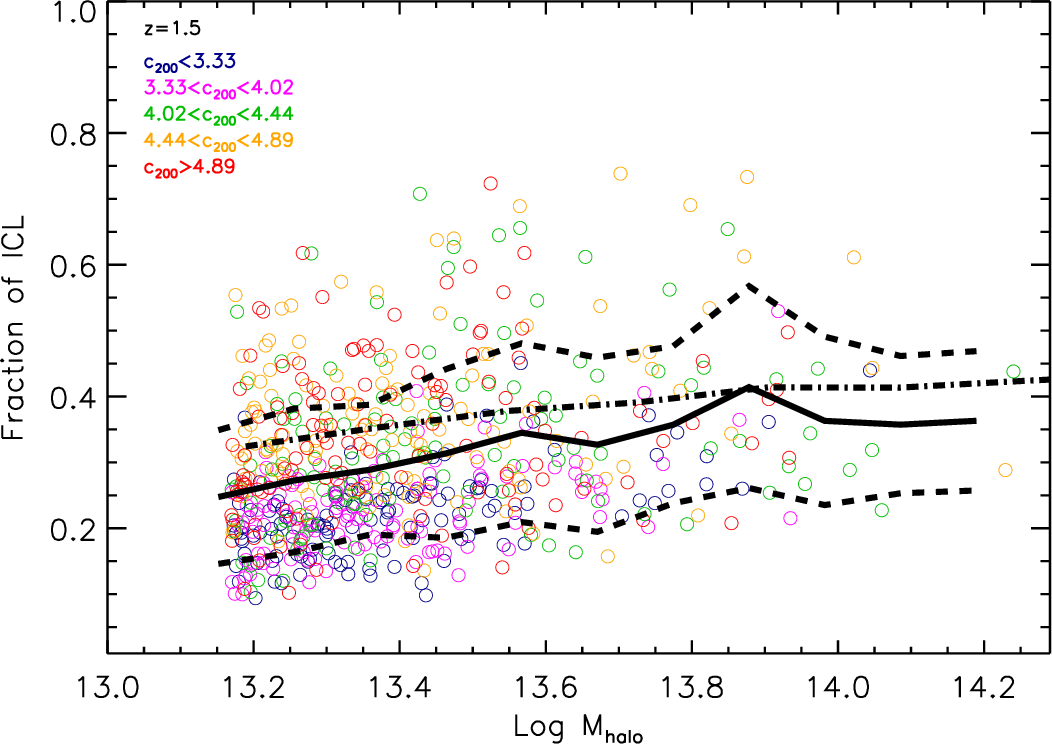} &
\includegraphics[scale=.48]{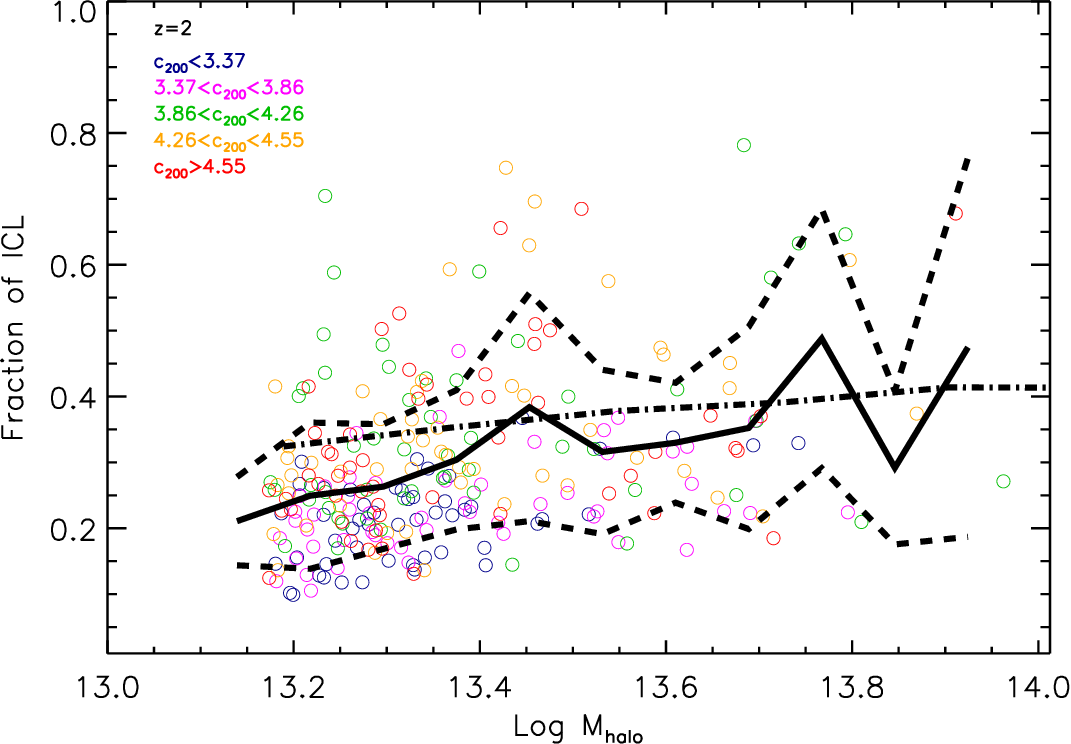} \\
\end{tabular}
\caption{Fraction of ICL as a function of halo mass, at different redshifts. The solid black line in each panel indicate the mean of the distribution, while the dashed lines represent $\pm 1\sigma$ scatter. In every panel, we also plot the relation at redshift zero represented by the dash-dotted line, while circles with different colors represent different concentrations from very high (red) to very low (blue). As found in C23a, the ICL fraction depends on the concentration of the halo, being higher for more concentrated objects. Here, the result found in C23a are extended to higher redshifts, up to $z=2$. Interestingly, there is no much evolution in the ICL fraction from $z=2$ down to the present time, which means that, overall, the ICL fraction is already important at high redshift, and does not evolve significantly with time, in agreement with recent observations.}
\label{fig:iclf_z}
\end{center}
\end{figure*}

\begin{figure*}
\centering
\includegraphics[width=0.73\textwidth]{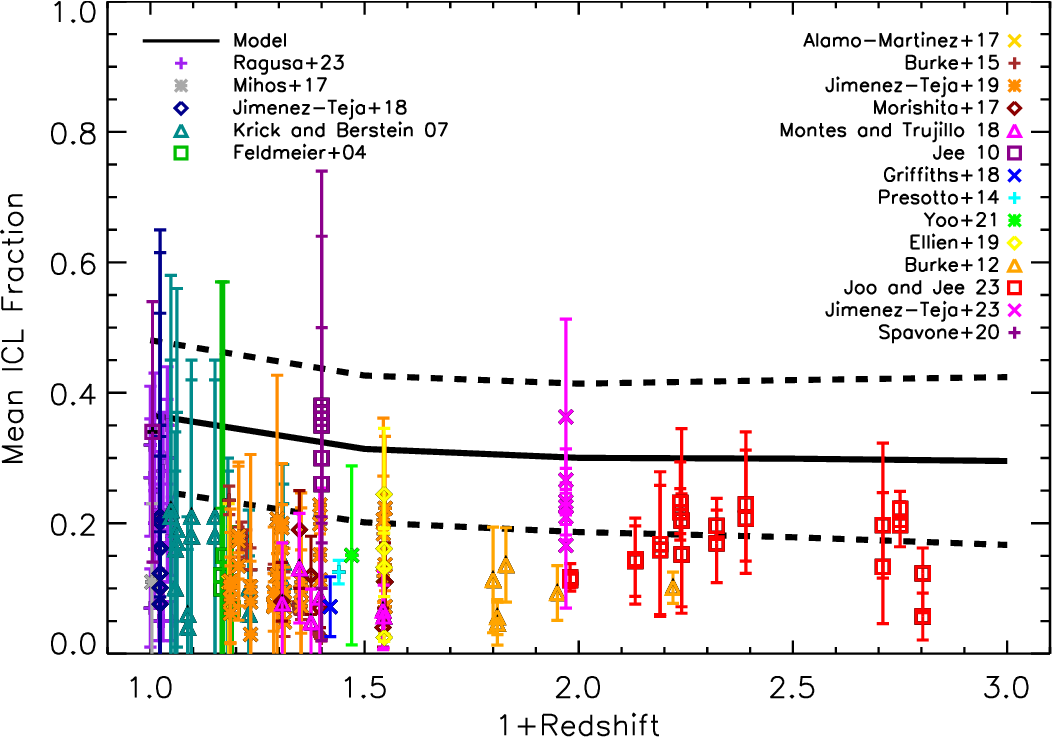}
\caption{Mean fraction of ICL as a function of redshift predicted by the model and its $\pm 1\sigma$ scatter, solid and dashed black lines respectively. Our predictions are compared with several
observations as indicated in the legend, spanning a range in redshift from $z=0$ to around $z=2$. It is important to mention that the observed data are taken within different apertures, while our fractions are taken within the virial radius $R_{200}$. Worth mentioning, although less important, is that the range of group/cluster mass is wide and different in both observations and our model at different redshift (see the text for more details). Considering the scatter, our model predictions represent fairly well what is observed, but it is quite clear that, overall, they are slightly higher than observations. The most important message of the plot is that our model confirms former observations (\citealt{joo2023}) that the ICL is already abundant at high redshift and its fraction is comparable with that observed in the local groups/clusters.}
\label{fig:ficl_z}
\end{figure*}

\subsection{ICL Fraction vs Redshift}
\label{sec:iclfrac}

Similarly to Figure 1 in C23a, in Figure 1 here, we plot the distributions of the ICL fraction at different redshifts, from $z=2$ (magenta) to the present time (black). Qualitatively speaking, the distributions at all redshifts can be approximated by a gaussian, which, as stated in C23a, is an indication of the fact that the processes responsible for the formation of the ICL are \textit{stochastic}. However, the main result of the plot is that the peak of the distribution tend to increase from higher to lower redshift, being around 0.23 at $z=2$ and reaching about 0.35 at $z=0$. This means that the mean ICL fraction gradually increases as time goes by, with a clear acceleration between $z=0.5$ (blue) and the present time. As such, the processes forming ICL become more and more important with decreasing redshift, but the bulk of the formation happens after $z=1$, and in particular between $z=0.5$ and the present time.

Figure 2 shows the ICL fraction as a function of halo mass at redshifts 0.5, 1, 1.5 and 2. For each panel, the mean ICL fraction and its standard deviation are represented by the solid and dashed black lines, respectively, while the dash-dotted black line indicates the mean ICL fraction at the present time. Circles with different colors represent different concentrations \footnote{The concentrantion is defined as in C23a, that is, by assuming an NFW (\citealt{nfw1997}) profile we derive the scale radius $R_s$ and then the concentration is calculated as $c_{200}=R_{200}/R_s$, where $R_{200}$ is the virial radius of the halo.}, in five bins from very high (red), to very low (blue), such that each bin contains the same number of objects. With this plot, we extend to higher redshift what was already found in C23a, i.e., that the ICL fraction depends very weakly on the halo mass. Most importantly, we extend the importance of the concentration in the ICL formation to higher redshift, i.e., more concentrated haloes tend to host a higher fraction of ICL. Another important feature shown by the figure can be seen by comparing, in each panel, the solid and dash-dotted lines. We can see that there is not much evolution in the ICL fraction between the highest redshift chosen and the present time, indicating that the ICL fraction is already important at high redshift and possibly comparable with that at $z=0$, in good agreement with recent observations (\citealt{ko2018,joo2023}).

In order to quantify the average ICL fraction at each redshift, we plot in Figure 3 the mean ICL fraction of the full samples of haloes as a function of redshift (solid and dashed black lines indicating mean and standard deviation), and compare the predictions of our model with a plethora of observed data from different authors (\citealt{ragusa2023,mihos2017,jimenez-teja2018,krick2007,feldmeier2004,alamo2017,burke2015,jimenez-teja2019,morishita2017,montes2018,jee2010,griffiths2018,presotto2014,yoo2021,ellien2019,burke2012,joo2023,jimenez-teja2023,spavone2020}).
There is no trend with redshift, which means that, as anticipated above, the ICL fraction assumes similar values independently of the redshift, with a mean and standard deviation of about $0.3\pm0.13$ that is just a little higher at $z=0$. At the present time, the mean ICL fraction is somewhat higher because we are sampling more massive haloes that tend to have a slightly higher fraction of ICL. This result is qualitatively in agreement with several observations, but quantitatively speaking, our model seems to be biased high with respect to them. There are at least two caveats worth mentioning and that are important in the context of the comparison between our model predictions and the observations. First, different authors compute the ICL fractions by using different apertures (see discussion in C23a). Second, the range in halo mass for which the ICL fractions are computed in observations spans a wide range as our sample does, but while in the former case they are single measurements, in our case is an average done over the whole sample for each redshift. Considering that the ICL fraction depends very weakly on the halo mass, the first caveat appears to be more relevant.

However, the interesting message coming from Figure 3 is that our model predicts the observed trend, accounting for similar ICL fractions even at high redshift, contrary to what several authors claimed so far about our model, i.e., that it predicts very low or even negligible fractions at  $z \gtrsim 1$.

\begin{figure}
\begin{center}
\begin{tabular}{cc}
\includegraphics[scale=.445]{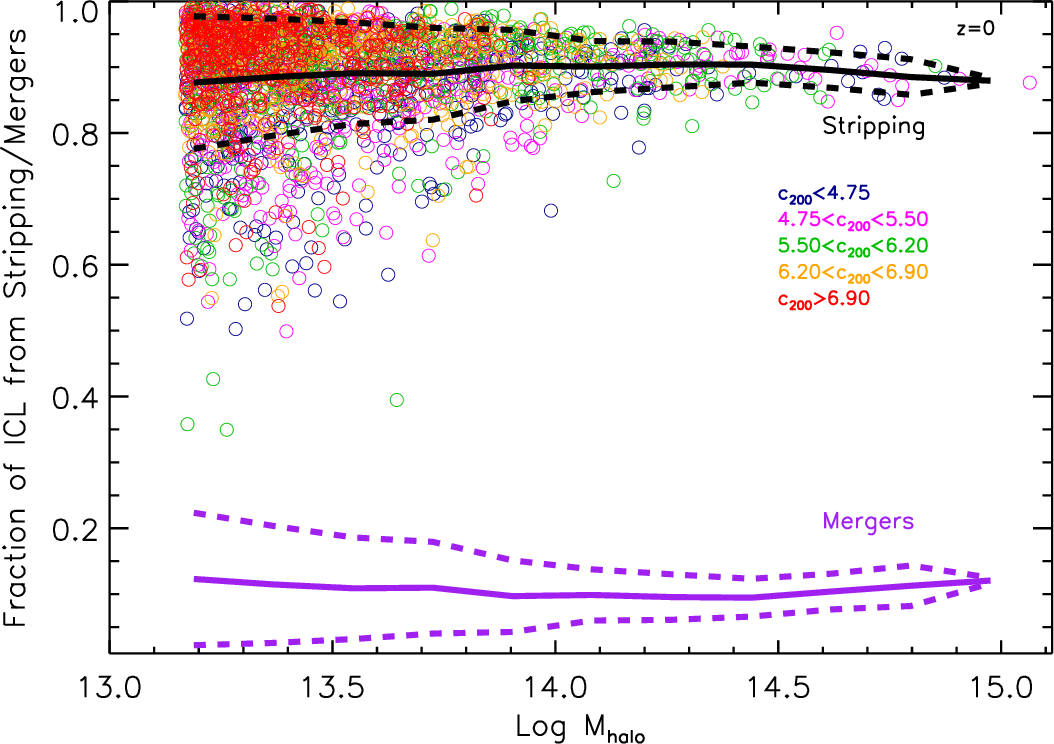} \\
\includegraphics[scale=.45]{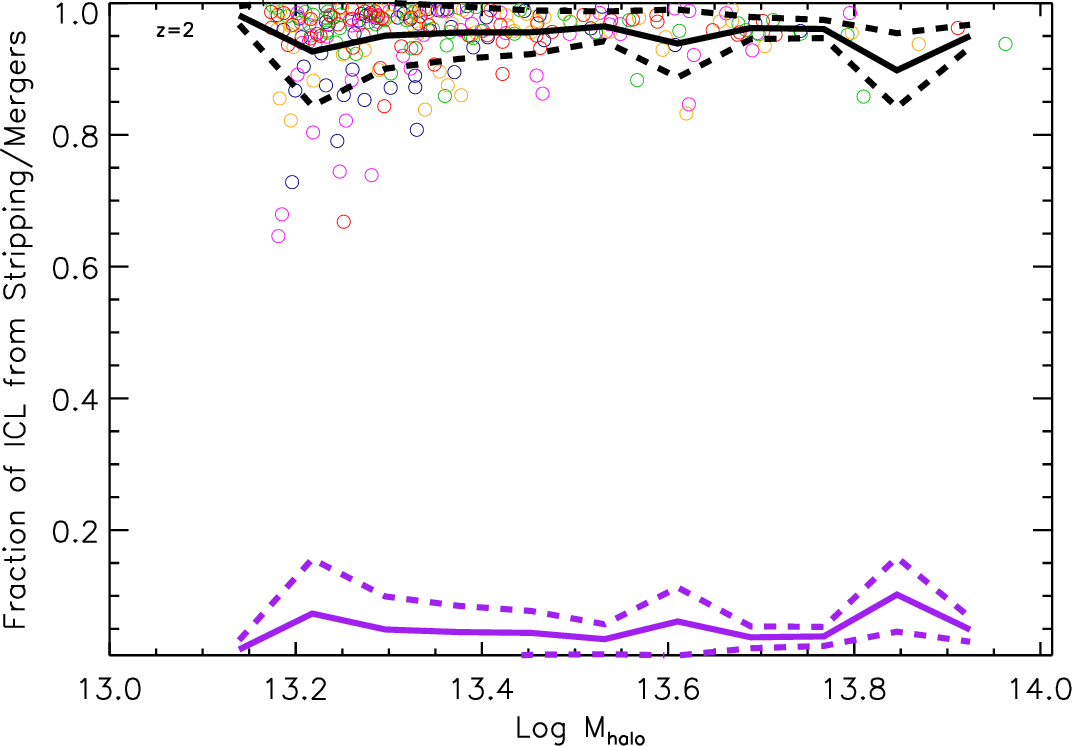} \\
\end{tabular}
\caption{Contribution to the ICL coming from the two direct channels for its formation, i.e., stellar stripping and mergers, as a function of halo mass. The upper panel refers to the model predictions at
$z=0$, while the bottom panel at the highest redshift probed in this work, $z=2$. Solid and dashed lines represent the mean and $\pm 1\sigma$ scatters, black in the case of stellar stripping and purple in the case of mergers, while the circles refer to single objects for the stellar stripping channel and are color coded based on the concentration as in previous figures. The main channel for the formation of the ICL is stellar stripping of satellites galaxies, independently of halo mass and redshift (for the sake of concision we omit the other redshifts, but the results are the same). On average, stellar stripping provides almost 90\% (including the amount from pre-processed ICL coming from this channel) of the total amount of ICL at the present time, and even slightly more at higher redshifts, leaving to mergers a marginal role with an average contribution of about 10\%  at most (present time). Importantly, the concentration has a remarkable role, as found in C23a. Indeed, stellar stripping is more efficient (higher fractions) in more concentrated haloes (red circles).}
\label{fig:stripmerg}
\end{center}
\end{figure}

\begin{figure*}
\begin{center}
\begin{tabular}{cc}
\includegraphics[scale=.48]{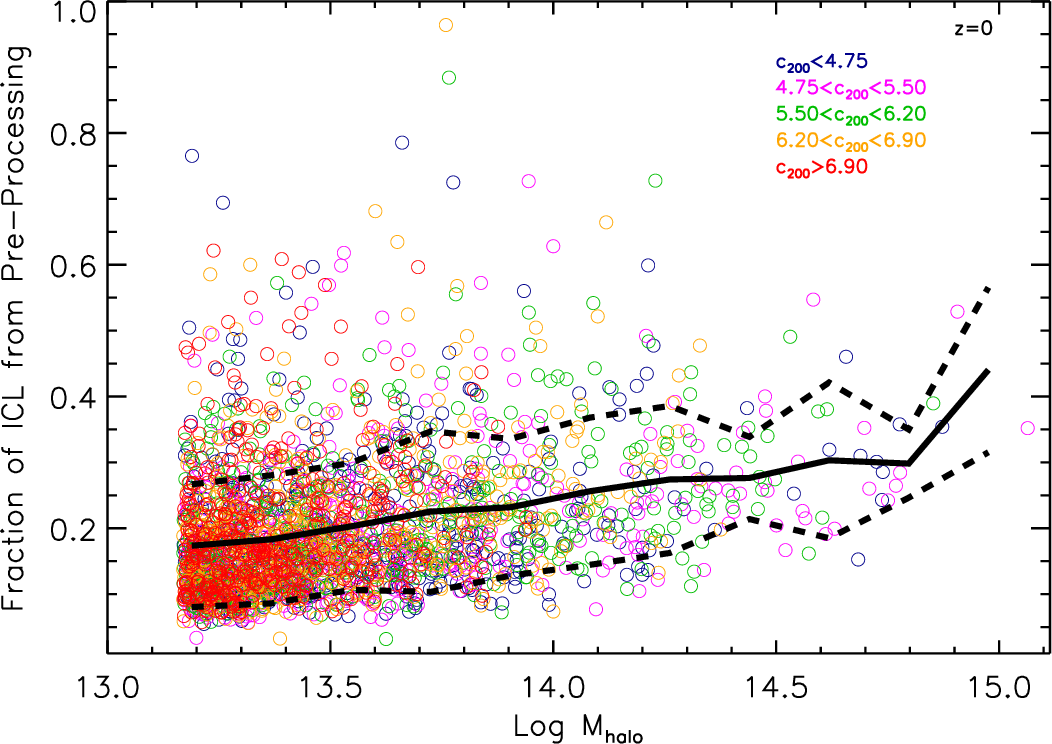} &
\includegraphics[scale=.48]{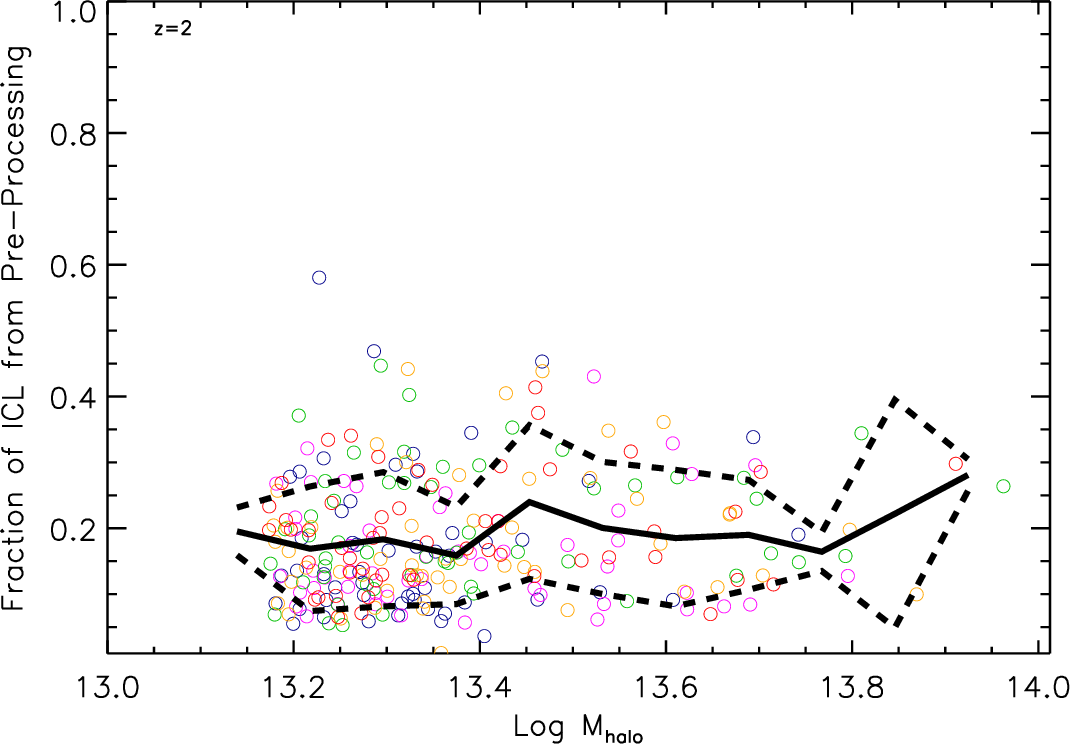} \\
\end{tabular}
\caption{Contribution to the ICL given by the only indirect channel for its formation, pre-processing or accretion, as a function of halo mass and redshift (different panels similarly to Figure 6). The solid black lines represent the mean while the dashed ones indicate $\pm 1\sigma$ scatters. As in previous figures, the circles with different colors represent haloes with
different concentrations. It appears clear that pre-processing or accretion of ICL from other haloes is an important channel to accumulate diffuse light, at any redshift (also in this case we omit the other redshifts). At the present time, the average fraction of ICL formed outside the halo and accreted during the process of its formation, can be as important as 20\% in low mass haloes, and up to 30-40\% in the largest ones. Moreover, although the scatter is around $\pm 10\%$, for a few cases the fraction of ICL formed ex-situ is higher than 50\%. At higher redshift the trend is similar and considering the scatter, the average fraction stays between 10\% and 40\%. Not surprising, the concentration does not have a relevant role, because the amount of ICL that can be accreted is supposed to be related to the accretion histories of single haloes rather than their central concentration. The only trend in concentration is horizontal, but it is due to the well-known relation that it has with the halo mass.}
\label{fig:accret}
\end{center}
\end{figure*}

\begin{figure}
\begin{center}
\begin{tabular}{cc}
\includegraphics[scale=.46]{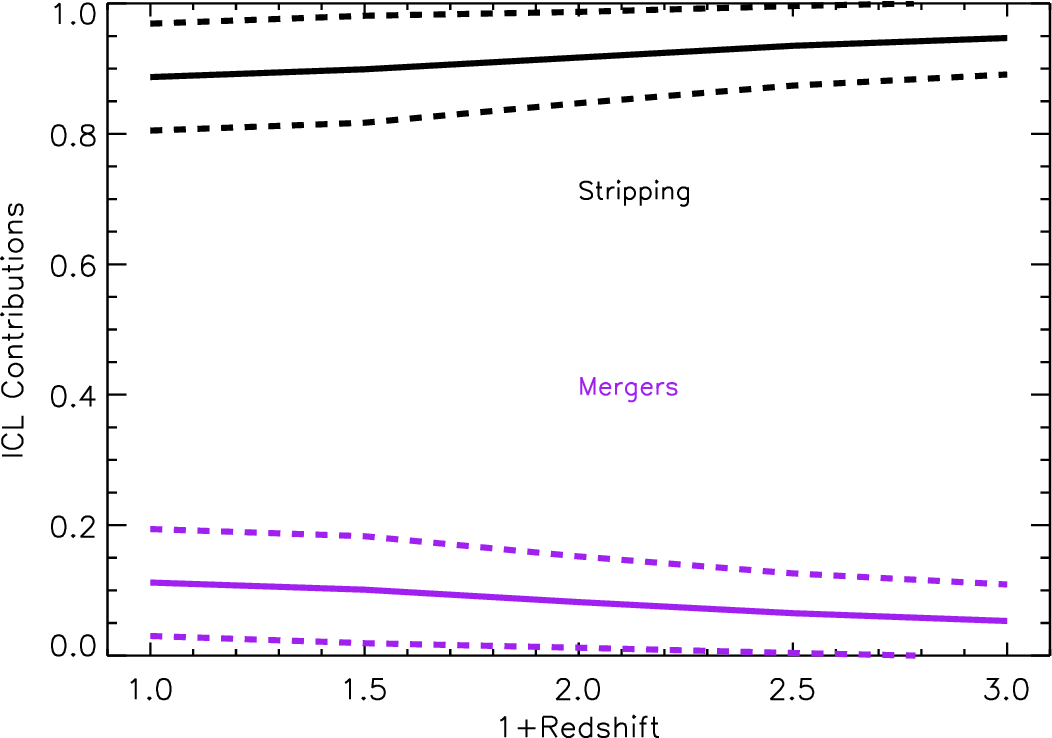} \\
\includegraphics[scale=.46]{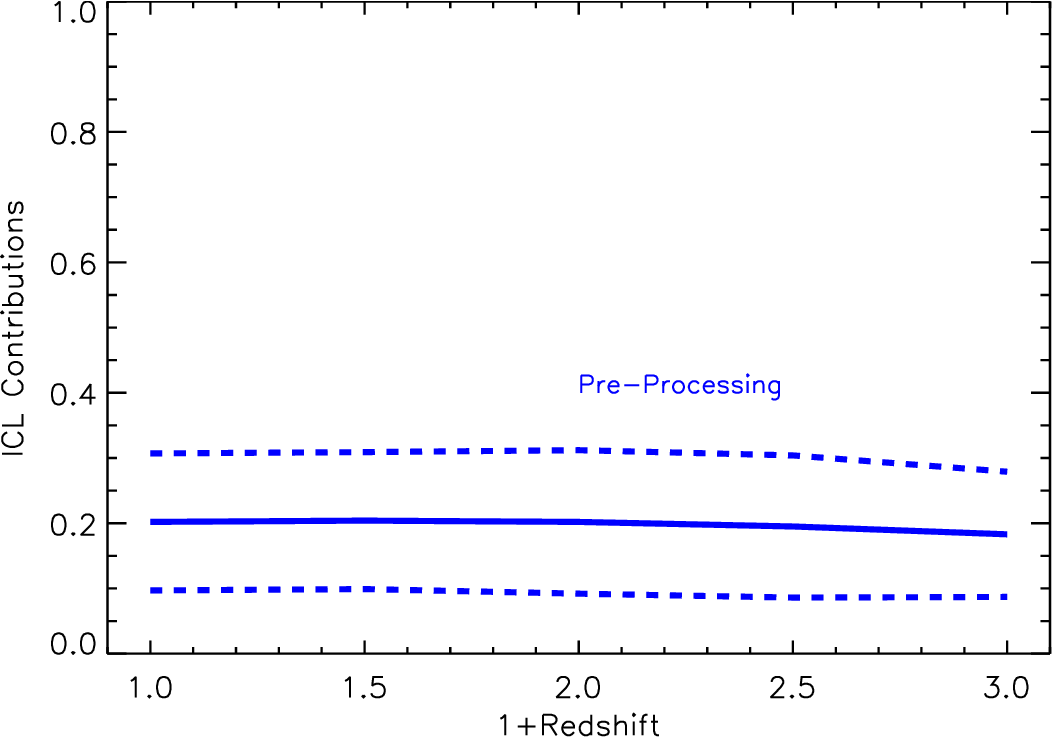} \\
\end{tabular}
\caption{Mean ICL contribution as a function of redshift given by the three channels: stellar stripping (black lines), mergers (purple lines) in the top panel, and pre-processing (blue lines) in the bottom panel. The solid lines refer to the mean fraction of all haloes at that redshift, while the dashed lines indicate $\pm 1\sigma$ scatter. As seen above, stellar stripping is by far the most important channel, while mergers give only a little contribution. Pre-processing, which we remind the reader to be an indirect channel, is anyway formed through either stellar stripping or mergers. This channel is more important than mergers between satellites and the BCGs, accounting for around 20\% of the ICL independently of the redshift. The net message of this plot is that none of the channels for the formation of the ICL has a preferred time during which their contribution is boosted or restrained.}
\label{fig:iclcontr_z}
\end{center}
\end{figure}

\subsection{Contributions from Different Channels}
\label{sec:contribution}

In our SAM, the ICL is formed via two direct processes, stellar stripping and mergers, but its mass can grow through the indirect pre-processing/accretion process. Our goal is to isolate the contribution given to the ICL by each of these processes, as a function of halo mass and redshift.

In Figure 4 we address part of the purpose, by showing the contribution to the ICL coming from the two direct channels, stellar stripping and mergers, as a function of halo mass and redshift ($z=0$ in the upper panel and $z=2$ in the bottom one). In both panels the solid and dashed lines represent the mean and the standard deviation (black for stellar stripping and purple for mergers), while circles indicate single objects for the stellar stripping channel, color coded based on the concentration of the host halo. It appears clear that stellar stripping dominates the formation of the ICL, at any halo mass and redshift probed (similar trends for redshifts in between). There is no trend with halo mass, which means that the efficiency of stellar stripping is independent on the mass of the host. On average, stellar stripping provides around 90\% of the total amount of ICL (including pre-processed ICL coming from this channel) formed until the present time, and slightly more at $z=2$. Mergers, on the other hand, have just a marginal role with an average percentage of about 10\% at most. Not surprising but important, the concentration has a remarkable role, in the sense that stellar stripping is more efficient, that is, higher fractions, in more concentrated haloes.

Stellar stripping and mergers account for the total of ICL formed but, as mentioned above several times, part of the ICL already formed through these two channels can be later accreted in larger objects,
and we refer to this process as pre-processing/accretion. Similarly to Figure 4, in Figure 5 we want to quantify the contribution that pre-processing provides, as a function of halo mass and redshift ($z=0$ in the left panel and $z=2$ in the right one). The solid and dashed black lines represent the mean and the standard deviation, respectively, while circles with different colors indicate single haloes with different concentrations. The pre-processing channel, i.e. accretion of ICL formed in other haloes, is an important way to accumulate ICL. In low mass haloes, the average percentage is around 20\%, but it increases to 30-40\% in large haloes. It is also worth noting that for a few haloes in our sample at the present time, more than half of the ICL has actually been accreted, and this happened even in haloes on large group scales ($\log M_{halo}\sim 14$). We find a similar trend at higher redshift, where the average fraction ranges from 10\% to 40\%. One more aspect to highlight regards the role of the concentration. In this case, the concentration is not playing any relevant role, and this is not surprising because the amount of ICL that can be accreted depends on the accretion histories of the haloes rather than on their central concentration.

To conclude the analysis on the contributions given by the different channels, in Figure 6 we plot the mean ICL contribution from each one of the channels (top panel for stellar stripping and mergers, and bottom panel for pre-processing), as a function of redshift. The solid and dashed lines represent the mean and the standard deviation, respectively, in black for stellar stripping, purple for mergers, and blue for pre-processing/accretion. Among the two direct channels (we remind the reader that pre-processing is an indirect channel), stellar stripping is the most important one at all redshift and by far, while mergers, as seen above, have just a marginal role. Pre-processing results in being more important than mergers, accounting for around 20\% of the ICL, independently of the redshift. The key feature of the plot is that there is no trend with redshift, implying that no channel has a preferred time to act, and so to be either boosted or restrained. We will come back on this point in Section 4 when we put the puzzle together.

\begin{figure}
\centering
\includegraphics[width=0.46\textwidth]{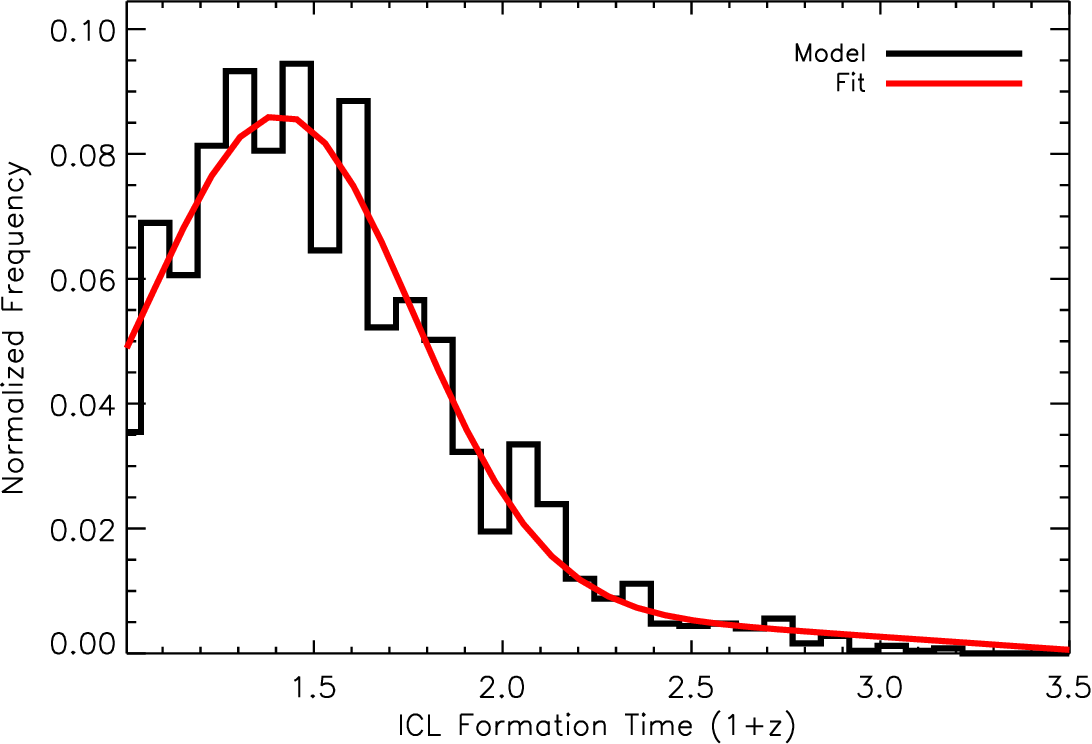}
\caption{Distribution of the ICL formation time, defined here as the redshift when 50\% of the ICL at the present time was already formed (see the text for a full discussion on the arbitrariness of the definition chosen). The black histogram indicates the distribution of the model predictions, while the red line refers to a Gaussian fit of the distribution. The ICL in the majority of our haloes forms very late, after $z\sim 1$, and the peak of the distribution is at around $z=0.5$. Clearly, the choice of a higher percentage in the definition of the ICL formation time would move the distribution towards lower redshifts, and vice versa for lower percentages. Interestingly, the plot shows also that there is a long tail populated by haloes for which their ICL forms early or very early, even at $z>2$.}
\label{fig:icltform_histo}
\end{figure}

\begin{figure*}
\begin{center}
\begin{tabular}{cccc}
\includegraphics[scale=.48]{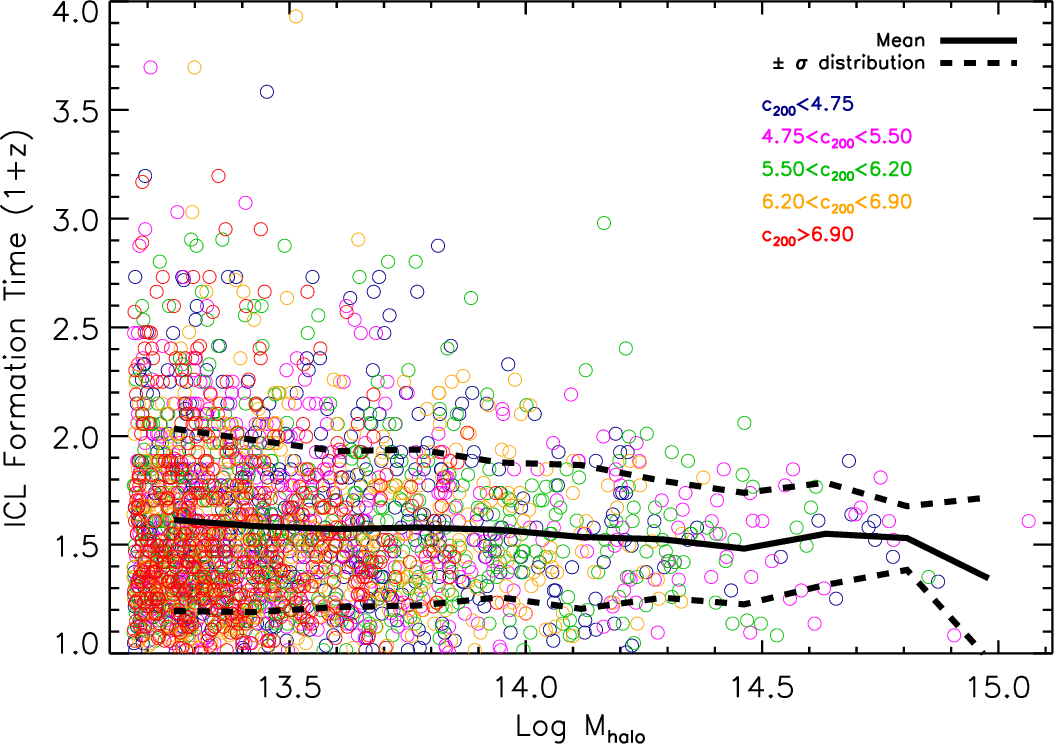} &
\includegraphics[scale=.48]{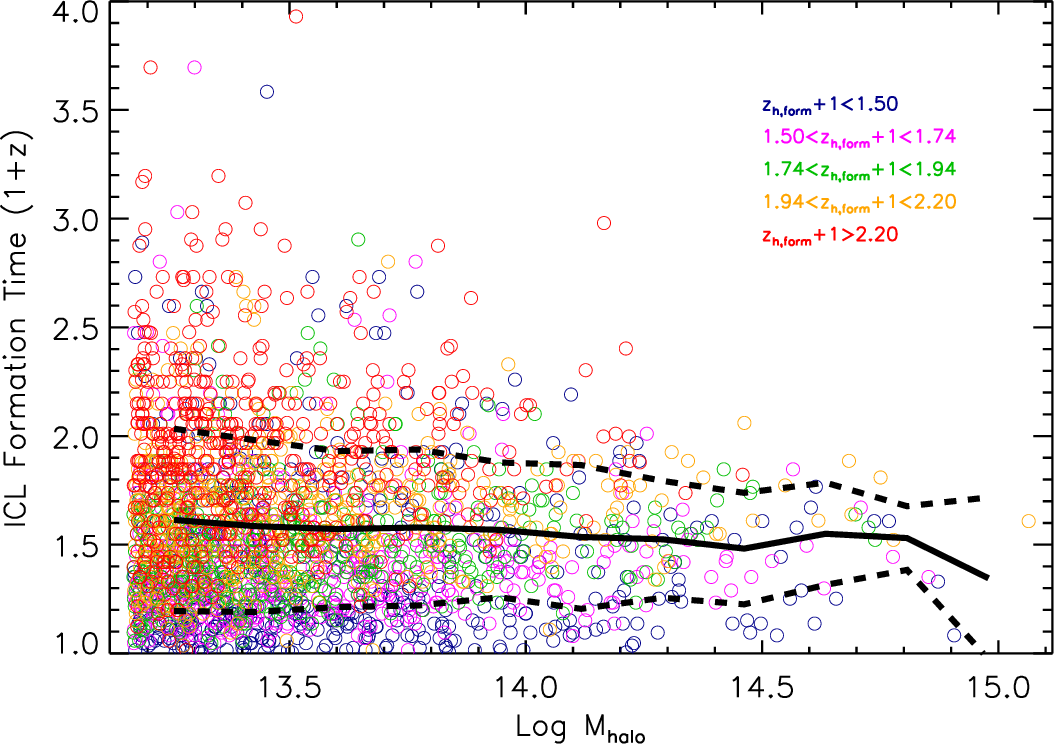} \\
\end{tabular}
\caption{ICL formation time as a function of halo mass, for haloes with different concentration (circles with different colors) at $z=0$ (left panel), and for haloes with different halo formation time (right panel). Similarly to the case of the ICL, here the halo formation time is defined as the redshift when the virial mass $M_{200}$ of the halo is 50\% of that at the present time. The solid black lines refer to the mean, and the dashed ones to $\pm 1\sigma$ scatters. There are three important aspects highlighted in this plot. The ICL formation time does not depend on the final mass of the halo, although the scatter (especially in very low mass haloes) is wide, and it depends on the concentration of the halo at the present time, considering the vertical trend in colors from redder to bluer ones. The most important trend is seen in the right panel, which clearly shows that in haloes formed earlier (redder colors) the ICL also forms earlier. This is evidence of the fact that the formation of the ICL is strictly connected to the formation history of the halo itself.}
\label{fig:icltform}
\end{center}
\end{figure*}

\begin{figure}
\begin{center}
\begin{tabular}{cccc}
\includegraphics[scale=.44]{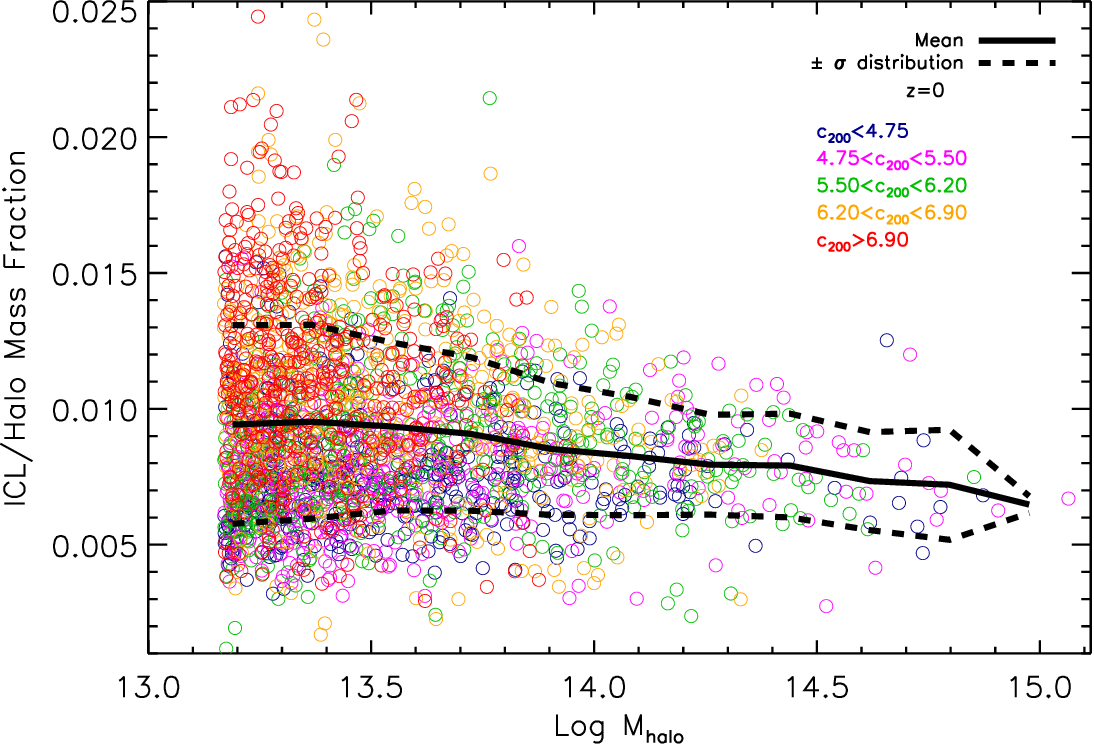} \\
\includegraphics[scale=.44]{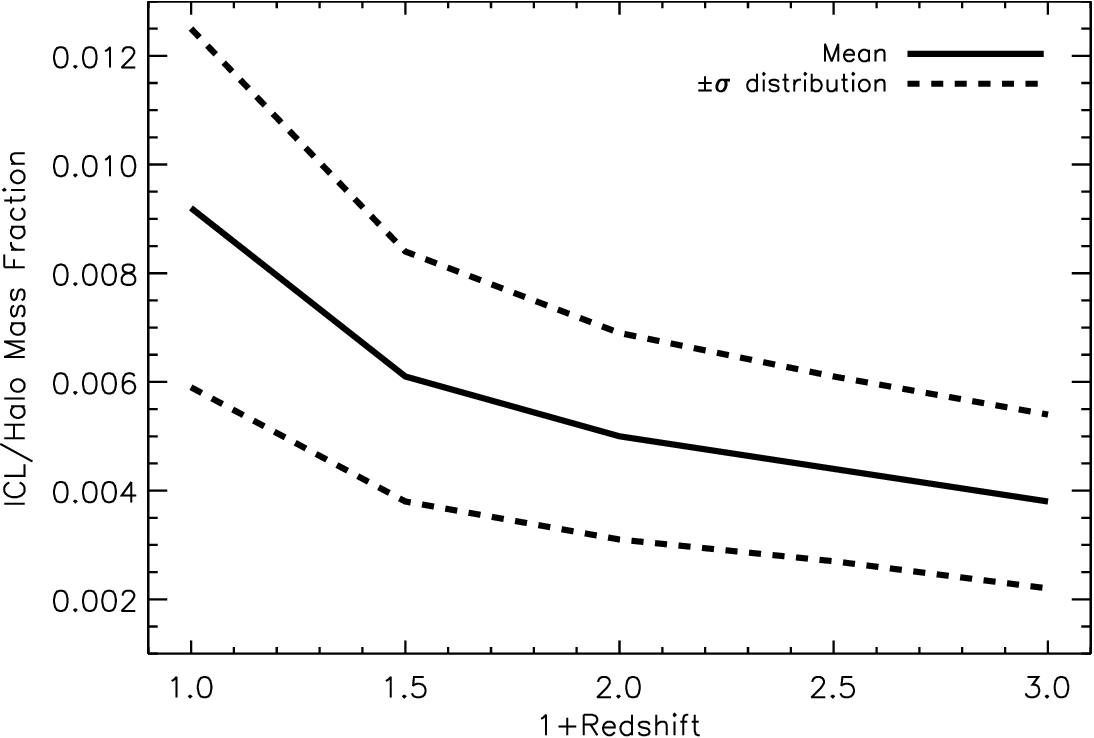} \\
\end{tabular}
\caption{Top panel: ratio between the ICL and halo mass as a function of halo mass at the present time. The circles with different colors represent haloes with different concentrations, while the solid and dashed black lines indicate the mean and $\pm 1\sigma$ scatter, respectively. The fraction in the y-axis can be considered as the efficiency of haloes in producing ICL. There is a weak trend, although the scatter is very large, for less massive haloes being slightly more efficient in producing ICL. Overall, by considering the scatter and the larger statistics in less massive haloes, the production of ICL is basically halo mass invariant, i.e., the main channels for the formation of the ICL produce a proportional amount of it independently of the mass of the halo. Once again, more concentrated haloes (redder colors) tend to be more efficient than less concentrated ones in producing ICL. Bottom panel: the same ratio plotted in the left panel, averaged among all haloes in the samples (solid and dashed lines indicate the mean and $\pm 1\sigma$ scatter, respectively), as a function of redshift. The average efficiency in forming ICL depends on redshift, being higher at lower redshift when the formation of the ICL is boosted. The difference between the present time and $z=2$ is more than a factor of 2.}
\label{fig:iclhalo}
\end{center}
\end{figure}

\subsection{Formation time and Efficiency of ICL Formation}
\label{sec:efficiency}

Considering that stellar stripping is the process responsible the most for the formation of the ICL, the amount or fractional amount of it should depend also on the available material to be stripped, i.e.,
on satellite galaxies that orbit close enough to the centre so that tidal forces can be important. On the other hand, the accretion of new members in the cluster depends on the assembly history of a given halo. Hence, the formation of the ICL should be somehow linked to the formation or assembly history of the host halo. Here, we want to analyze this aspect.

In Figure 7 we show the distribution of the ICL formation time, defined as the redshift at which 50\% of the ICL at the present time was already formed. For the sake of simplicity, the amount of ICL coming from pre-processing is considered formed when it is accreted in the main progenitor branch of the final halo. The model prediction is indicated by the black histogram, while the red line represent a gaussian fit of the distribution. We caution the reader that the shape of the distribution does not change if we choose a smaller or higher percentage in the definition of the ICL formation time, although the peak of it moves towards the right/left for smaller/higher percentages. Our choice is quite arbitrary, and we decided to use
50\% simply because it is the most common choice when defining the halo formation time (see discussion below). The distribution shows that the ICL in the majority of our haloes forms very late, after $z=1$ as found in \cite{contini2014}, and the peak of the distribution is at around $z=0.5$. However, the distribution has a long right tail, implying that for a non-negligible number of haloes the ICL can form early or even very early, such as $z>1$.

We use our definition of ICL formation time to analyze it as a function of halo mass, in order to see whether or not the mass of the host has a relevant role. The results are shown in Figure 8, where the solid and dashed black lines represent the mean and standard deviation, respectively. In the left panel, the circles with different colors indicate haloes with different concentrations, while in the right panel the circles indicate haloes grouped in different bins of halo formation time. As for the ICL, the halo formation time is defined here as the redshift when half of the virial mass $M_{200}$ of the halo at $z=0$ is formed. The two panel together show three important aspects. First, the final halo mass is not a good tracer of the ICL formation time, given the fact that from low to high mass haloes the typical ICL formation redshift is around 0.5-0.6. Second, although not clear as in former figures, the concentration of the halo at the present time correlates with the ICL formation time. This is more likely an inheritance from the main progenitor of the final halo, in the sense that if the halo is more concentrated than others of the same mass, its progenitors likely were in the past. We will further discuss this point in Section 4. Third, and most important, the ICL formation time strongly depends on the halo formation time (right panel, circles with different colors), in a way that for halo formed earlier, the ICL formed earlier too. This is clear evidence of the fact that the two formation histories go in parallel, that is, the formation of the ICL is closely connected to the formation history of its host. Also, the relation between the ICL and halo formation times is consistent with stripping being the main channel of the formation. Indeed, in haloes that form earlier, their member galaxies have more time to lose stellar mass via tidal stripping, and so resulting in a larger ICL fraction.

In C23a we concluded that the fraction of ICL at $z=0$ is weakly dependent on the halo mass, which basically means that the processes responsible for the formation of the ICL are halo mass invariant. Here, we extended the same result up to $z=2$. However, the fact that the amount of ICL in haloes is halo mass invariant does not necessarily means that the efficiency in producing ICL is the same at any
time, for a given halo mass. To conclude the analysis, we now want to focus on this aspect, by looking at the "efficiency" of ICL formation for haloes of different mass and at different redshifts. In the top panel of Figure 9 we plot the ratio between the ICL and halo masses as a function of halo mass at $z=0$. Solid and dashed black lines represent the mean and standard deviation, while circles with different colors represent haloes with different concentrations. This ratio can be considered as a sort of "efficiency" in producing ICL. The mean values range from about 0.009 in low mass haloes, to 0.006 in large haloes, but considering the scatter in the relation is possible to state that there is no clear trend between the two quantities. This means that, as already found previously in C23a, on average stripping and mergers together produce a proportional amount of ICL independently of the mass of the halo. However, once we plot the same ratio averaged among all haloes as a function of redshift (bottom panel of Figure 9), the average efficiency in forming ICL becomes higher with decreasing redshift (solid and black lines represent mean and standard deviation, respectively). The net increase of the efficiency between $z=2$ and $z=0$ is a factor of 2. If we consider the gap in time when the ICL formation is boosted (between $z=0.5$ and $z=0$), the increase in efficiency is a factor of 1.5, i.e., on average haloes at the present time are 50\% more efficient in producing ICL than their $z=0.5$ counterparts.

In the next section we will discuss the analysis done and the consequent results, by giving particular attention to their implication in the context of the formation of the ICL and its link with the assembly of dark matter haloes.

\section{Discussion}
\label{sec:discussion}

In the last years many authors focused on the link between the ICL and the dynamical state of its host (see \citealt{contini2021} for a review), by analyzing different aspects that could improve our knowledge about the distribution of the dark matter in haloes. All the studies have in common the use of the ICL as a beacon for the dark matter. It appears then clear that the ICL can be used as a luminous tracer of some properties of haloes in which it resides, but it is also fundamental to know its formation and assembly histories in order to differentiate the single cases. For instance, a halo that have recently experienced a merger has also a high probability that the distributions in colors and/or metallicity of the ICL would not show any gradient, while a more relaxed object would, due to the particular way the ICL forms. Below, we discuss the most relevant aspects of our analysis also in the light of former observational and theoretical findings.

\subsection{The ICL Budget in Groups and Clusters}
\label{sec:budget}

In Section 3.1 we have analyzed the ICL budget in groups and clusters as a function of redshift. One important result from our analysis is that the model predicts similar fractions as observed in the local universe, meaning that the ICL is already an important component at high redshifts. This conclusion is in perfect agreement with recent observations of objects at medium/high redshifts (e.g., \citealt{ko2018,joo2023,jimenez-teja2023}). Very recently, \cite{joo2023} studied the ICL fraction in ten clusters in the redshift range $1<z<1.8$ based on deep infrared imaging data, and found fractions very similar to those that are observed at lower redshift, with an average fraction of about 17\%. They did not find any significant trend with the halo mass or any colour gradient, results that they interpreted as the ICL forming in parallel with the formation of the BCG, through mergers and accretion of pre-processed material. The point regarding the channels suggested in their work will be discussed below, but here it is important to mention two interesting aspects: their observed fractions of ICL are similar to what our model predicts (although we remind the reader that our model is slightly biased high with respect to them), and that they do not see any colour gradient. The lack of a gradient in colors points to mergers as the main responsible for the amount of ICL formed so far, which is in line with a picture where the ICL follows the BCG formation at high redshift. As discussed in \cite{jimenez-teja2023}, mergers can be the primary channel at higher redshift and stripping at lower redshift, the so-called two-stage scenario of the ICL formation. These authors measured the ICL fraction of the {\small SPT-CLJ0615-5746} cluster at $z=0.97$ using its RELICS (Reionization Lensing Cluster Survey) observations in the optical and infrared. Starting with the hypothesis that the ICL fraction is a good tracer of the dynamical stage of an object, they found fractions ranging from 16\% to 37\% (depending on the particular wavelength), which they interpreted as a signature of merging, in contrast with former studies (e.g., \citealt{morandi2015,bartalucci2017}) that claimed the relaxed stage of the object in question. Regardless the dynamical stage of this cluster, the fractions are close to those found at much lower redshift, so putting another data point on the list of objects that prove the absence of trend with redshift, in agreement with the predictions of our model.

The key-point of our analysis is not only that the model predicts the observed trend with redshift and similar values, but also the role played by the concentration. Indeed, in C23a we showed that the concentration of the halo plays a major role in the formation of the ICL, in a way that the ICL fraction is higher for haloes that are more concentrated. More concentrated haloes means also more relaxed, and this conclusion goes in contrast with the interpretation given by \cite{jimenez-teja2023}, i.e., that higher fractions are expected in merging systems rather then in relaxed and passive ones. Our interpretations of the role played by the concentration has a pure dynamical reason: the higher the concentration, the stronger the tidal forces acting on satellite galaxies that provide the stray stars of the ICL. In this work, we extend the result to higher redshift, meaning that the concentration of a given halo at a given redshift makes a relevant difference in the ICL budget.

\subsection{Main Channels of the ICL Formation}
\label{sec:channels}

Several mechanisms for the ICL formation have been suggested in the past years, but still nowadays there is not a general consensus on which one/ones are the most important in terms of mass that it is
effectively moved to the ICL component through them. Even less consensus is achieved when it comes to the relative contribution of different mechanisms as a function of time, being one or another more important at a given range of time rather than the same mechanism at all times. Among the most invoked mechanisms, as briefly mentioned in Section 1, we find stellar stripping, mergers and pre-processing/accretion. For the sake of completeness, also dispruption of dwarf galaxies (e.g. \citealt{purcell2007,raj2020}) and in situ star formation (\citealt{puchwein2010}) have been proposed, but later discarded by several authors for different reasons. Disruption of dwarfs appears to be relevant in terms of number of galaxies that are destroyed during the process of the ICL formation, but it does not in terms of actual stellar mass that becomes unbound (\citealt{contini2014}). On the other hand, in situ star formation has been ruled out by observations (\citealt{melnick2012}) given the fact that its contribution is very little, around 1\%.

In our model the ICL forms via the three channels mentioned above, but they have different relative importance. We have seen in Section 3.2 that stellar stripping is the major channel for the ICL formation, and it dominates at every halo mass and redshift investigated. Mergers, on the other hand, have just a marginal impact that stays within 10\% regardless the halo mass and the redshift. Pre-processing appears to be more important than mergers, but way less than stellar stripping. This result is particularly in accordance with observational studies (e.g., \citealt{joo2023,proctor2023}) that give to pre-processing/accretion a relevant role in the ICL formation. However, as mentioned in this work and discussed in C23a, this channel is not a direct way to form ICL, because the amount of stray stars coming from pre-processing have already formed elsewhere in the past, and they are just accreted during the hierarchical formation of a given halo. This concept is fundamental when studying the ICL, simply because the properties of this part of the ICL would depend on the general properties of galaxies contributing to it in the original halo where it was formed, rather than on the properties of the halo and its galaxies where it was finally accreted.

Another key-aspect of our analysis regards the efficiency of stellar stripping and mergers as a function of halo mass and redshift. Hence, the efficiencies of these processes are halo mass and redshift invariant, which means that there is not a characteristic halo mass or a particular time where and when one of these two processes contribute more to the total ICL. Conversely, the pre-processed ICL depends weakly on halo mass, which is reasonable considering that more massive haloes are expected to accrete more material during their growth, but again it does not depend on redshift.

The picture coming from the analysis done here, together with former studies done using the same model is the following: the bulk of the ICL forms mainly after $z\sim 1$ via stellar stripping of intermediate/massive galaxies orbiting around the BCGs, while just a small amount comes from mergers between the satellites and the central galaxies. On the total of the ICL, which is formed through these two channels, a percentages that goes from 20\% (low mass haloes) to 40\% (high mass haloes) is accreted from smaller objects during the formation of the final halo itself. Many observational studies have given a relevant role to stellar stripping (e.g., \citealt{demaio2015,demaio2018,montes2018}), but also others stated the importance of mergers. For instance \cite{groenewald2017}, assuming a relative contribution of 50\% from major mergers would explain the stellar mass growth of the ICL from $z=0.3$ to the present time. \cite{jimenez-teja2023} (but see also \citealt{jimenez-teja2018}) found a link between ICL and dynamical state of a cluster, in the sense that objects that have recently experienced a merger are also those with higher budget of ICL, so giving a particular relevance to this channel.

On this context, it is worth mentioning two remarkable aspects. First and foremost, also our analysis shows that mergers can be relevant in a few cases, in particular for low mass haloes, although on average, it is stellar stripping to account for the most of the ICL. There might be cases, at all redshifts, where a single major merger or multiple minor mergers are responsible for the bulk of the ICL formed. It does not happen with the same frequency on cluster scale given the much higher number of satellites that can be subject to stripping. Second, as discussed in previous studies (see, e.g. \citealt{contini2018}), the definition of a merger is not unique, which means that part of what is called stellar stripping here can actually be from the merger channel in other works. A clear example of it is the parallelism made in \cite{contini2018} between the results found there and those found by \cite{murante2007}. In that study we found similar contribution coming from stellar stripping in the innermost central regions of a group/cluster, in contrast with \cite{murante2007} who found that around 75\% of the ICL comes from the merger channel. However, when we relaxed our boundary between stellar stripping and merger channels, we found a similar fraction in perfect agreement with that claimed by \cite{murante2007}.

\subsection{Link between the ICL and Its Host Halo}
\label{sec:efficiency2}

We have seen that stellar stripping dominates the formation of the ICL at all halo masses and redshifts, but a substantial amount of already formed ICL comes from the accretion of external material (pre-processing channel). As mentioned in Section 3.3, it is easy to assume that the growth of the ICL depends on the growth of the host halo, in the sense that the more the halo grows, the likely the ICL does. This can happen because of two reasons: on one hand, a sustained accretion history of the host halo would bring more ICL coming from the pre-processing channel and, on the other hand, it would also bring more satellite galaxies that can be subject to stellar stripping. Under this point of view, it appears clear that the accretion history of the host can be extremely important in the final budget of the ICL, thus linking the final dynamical stage of the halo with the amount of ICL in it.

A way to investigate the link between ICL and its host is by looking at the typical time of their formation and see whether or not the two quantities are somehow connected. We have shown in Figure 7 the distribution of the ICL formation redshift, and concluded that even though in most of the objects in our sample the ICL forms late, after $z=0.5$, there is a long right tail in the distribution which indicates that for these objects the ICL formed very early, even before $z=1$.  These host haloes have had early growth that enhanced the growth of the ICL in early stages, while for the most of haloes in our sample, a smoother growth of the haloes reflected on smoother growth of the ICL too.

What makes the difference in the ICL formation time? In order to answer to this question, in our analysis we have considered three quantities that could be related to the ICL formation time, namely: the final concentration, the mass and formation time of the host halo. Both ICL and host formation time have been defined as the redshift when half of the ICL/halo mass at the present time was already formed (we remind the reader that qualitatively speaking results do not change by choosing lower or higher percentages). We found that the ICL formation time does not depend on the final mass of the host halo, meaning that it cannot be used as a tracer of the typical ICL formation time. On the contrary, there is a dependence of the ICL formation time with the final concentration of the host halo, which we claim to be a natural consequence of the fact that if a halo, at fixed mass, is more concentrated at the present time, it would have likely been in the past. If so, the formation of the ICL for more concentrated haloes is smoother than those less concentrated, bringing to the net result of later formation times for the former.

A different picture comes out when we compare the ICL and halo formation times. The two quantities are strictly connected: earlier host formation time means also earlier ICL formation time, clear evidence that ICL and its host growths go in parallel. What really makes the difference in the formation time of the ICL is the growth history of its host. Considering all the results so far mentioned, the final mass of a given halo is not important for the final budget of the ICL. However, at fixed halo mass, more concentrated haloes, and so haloes of earlier formation, form more ICL since early epochs. The net result at the present time is that haloes with a higher fraction of ICL are also haloes formed earlier and that are, and probably have been, more concentrated than those having a smaller fraction of ICL. Here is the intimate connection between the ICL and its host. Hosts more concentrated are also more relaxed, which means that the ICL is more abundant (proportionally to the total stellar mass in the host) in objects that are in the final stage of their growth.

The channels responsible for the formation of the ICL have then been enhanced in hosts where the ICL fraction is higher than the average. This also means that the efficiency in forming ICL in these objects is higher. We have seen that the fractional amount of ICL is invariant with respect to the halo mass, but it is higher in those that are more concentrated. In the top panel of Figure 9 we have seen that the ICL over halo mass ratio, which can be considered as a sort of efficiency in forming ICL, does not depend on halo mass in a clear way, but it does when haloes are selected based on their concentration. The efficiency is higher for haloes that are more concentrated, which highlights one more time the remarkable role played by the concentration in the context of ICL formation. Although the efficiency is mass invariant, there is a net dependence on redshift, being higher as time goes by. The reason for this trend lies on the fact that the concentration is the quantity playing the relevant role. Indeed, more concentrated haloes at the present time more likely come from progenitors that were more concentrated at high redshifts. However, regardless of the mass, at high redshift haloes are less concentrated and dynamically younger, or less relaxed (e.g., \citealt{gao2011,contini2012}), which means that the efficiency in forming ICL is enhanced with time because the concentration acquires more and more importance as it increases with time, and the concentration is the quantity responsible for the efficiency of the most important channel, that is, stellar stripping.

To summarize, the bulk of the ICL is formed via stellar stripping, a mechanism that, at fixed mass, is more efficient in haloes that are more concentrated, relaxed and that formed earlier than their counterparts that are less concentrated, less relaxed and that formed later. This conclusion, as anticipated above, is in contrast with the two-stage theory of the ICL formation invoked by some authors (e.g., \citealt{jimenez-teja2023} and references therein), where the ICL forms mainly through mergers at $z>1$, and through stripping of satellites galaxies at later epochs. Not only, it is in contrast with the recent results of \cite{jimenez-teja2023} who found a higher fraction of ICL in clusters that recently experienced a major merger, with respect to those that are already relaxed and so, in an advanced dynamical stage. Nevertheless, our findings are in agreement with the observed trend by \cite{ragusa2023}, i.e., a higher fraction of ICL in haloes having also a higher fraction of early-type galaxies, considered to be a proxy of the dynamical stage of a given halo (higher fractions in more evolved, relaxed haloes).

Although the first attempts to link the ICL with the dynamical state of the halo are more than a decade old by now (see, e.g., \citealt{zibetti2005,jee2010,giallongo2015}), it has been only in the very recent few years that the ICL community
has started to extensively focus  on the topic, theoretically (\citealt{harris2017,pillepich2018,asensio2020,contini2020b,deason2021,yoo2022,chun2023}) and observationally (\citealt{montes2018,montes2019,zhang2019,kluge2021,sampaio2021}).  \cite{harris2017} have been the first authors to conclude, in a theoretical way, that the ICL is more centrally concentrated than the DM, a result which the following works by \cite{contini2020b,contini2021b} and \cite{contini2022} relied on in linking the concentration of the ICL distribution with that of the DM, with the assumption of an NFW profile. From the observational side, \cite{montes2019} were the first to quantify the typical average distance between the ICL and DM distributions, result qualitatively and quantitatively confirmed in the theoretical analysis by \cite{asensio2020} and \cite{yoo2022}. In particular, the latter found that among the sample of haloes they simulated, the relaxed ones show a closer similarity in the ICL+BCG and DM distributions with respect to those that are dynamically younger, a conclusion that is very close to the results presented here and in C23a. \cite{kluge2021} focused on the links between BCG and ICL alignments with their host cluster, and concluded that the ICL aligns with its host better than the BCG in terms of position angle, ellipticity and centering. Even more importantly in the context of the current work, they confirmed the prediction of \cite{harris2017}, i.e., that the ICL is more concentrated than the DM. 

Given the importance of the topic, and the different conclusions that came out from independent analysis so far, it is necessary to probe more the link between an observable quantity such as the ICL and the properties of its host halo, both observationally and theoretically, in order to learn more about the dynamical state of galaxy groups and clusters, especially for single case studies.

\section{Conclusions}
\label{sec:conclusions}

We took advantage of a set of dark matter only simulations run with {GADGET4} code, that provided the merger trees on which our state-of-the-art semi-analytic model has been run, to study the ICL fraction, formation time and the contributions given by the channel responsible for its formation, as a function of halo mass, redshift and concentration of the host halo. Based on our analysis, done on a wide sample of haloes at every redshift probed, and on the discussion of the results, our main conclusions can be summarized as follows:
\begin{itemize}
	\item We extend to higher redshifts, $z=2$, the result found in \cite{contini2023a}, where the ICL fraction is weakly dependent on halo mass and strongly dependent on the concentration of the host halo, being higher for more concentrated objects. Moreover, the predicted ICL fraction at higher redshift is comparable with that found at the present time, in good agreement with recent observational measurements;
	\item The ICL forms mainly via three channels: stellar stripping and mergers (direct channels), pre-processing/accretion (indirect channel). Stellar stripping is the most important process for the ICL formation, considering that it accounts for around 90\% of the total ICL budget, leaving to mergers just a marginal role (around 10\%), regardless of halo mass and at any redshift. Pre-processing, i.e., ICL stars that formed outside the host halo and that have been later accreted, explains about 20\% of the ICL in low mass haloes, and up to 40\% on cluster scales at the present time, with similar percentages at higher redshifts; 
	\item The ICL forms very early, after $z\sim 1$ with a peak at $z=0.5$. However, a considerable number of haloes in our sample have formed their ICL much earlier, up to $z\sim 2$. More importantly, the ICL formation time depends both on the concentration and formation time of the host. More concentrated and early formed haloes form their ICL earlier than less concentrated and later formed ones. We interpret this result as follows: in haloes that form earlier their satellites have more time to be subject to stellar stripping and lose their stellar mass, a process that becomes progressively important when they reach the central regions where, given the higher concentration, tidal forces are stronger;
	\item The efficiency of the ICL formation is halo mass invariant, but it strongly increases with decreasing redshift. This is the net result of the role played by the concentration. At any time, the concentration is the quantity regulating the power of stellar stripping (the most effective channel). The difference in efficiency as a function of time is due to the fact that haloes of the same mass at high redshift are less concentrated and dynamically younger, i.e., stellar stripping becomes more and more efficient as time goes by, but its efficiency is halo mass independent;
	\item At fixed halo mass, more concentrated, early formed and more relaxed objects have a higher fraction of ICL. This links the ICL to the dynamical state of its host, in a way that a group or cluster with a higher fraction of ICL than the average is in a more advanced stage of growth.
\end{itemize}
In a forthcoming study, the last of the series, we will address the same points addressed in \cite{contini2023a} and here, but probing much lower halo mass scales, down to Milky-Way like haloes. Our main goal will be that of understanding if the results we have found so far and their explanations can be found and applied to lower scales.


\section*{Acknowledgements}
The authors thank the anonymous referee for his/her constructive comments which helped to improve the manuscript, and Hyungjin Joo for providing the observed data in Figure 3.
E.C. and S.K.Y. acknowledge support from the Korean National Research Foundation (2020R1A2C3003769). E.C. and S.J. acknowledge support from the Korean National Research Foundation (RS-2023-00241934). All the authors are supported by the Korean National Research Foundation (2022R1A6A1A03053472). J.R. was supported by the KASI-Yonsei Postdoctoral Fellowship and was supported by the Korea Astronomy and Space Science Institute under the R\&D program (Project No. 2023-1-830-00), supervised by the Ministry of Science and ICT.


\bibliography{paper}{}
\bibliographystyle{aasjournal}

\end{document}